\documentclass[aps,prb,graphicx,twocolumn]{revtex4-1}

\usepackage{graphicx}
\usepackage{dcolumn}
\usepackage{bm}

\usepackage[utf8]{inputenc}
\usepackage[T1]{fontenc}
\usepackage{mathptmx}
\usepackage{amsmath}

\begin{document}


\title{Optimized design of a Silica encapsulated photonic crystal nanobeam cavity for integrated Silicon-based nonlinear and quantum photonics } 



\author{J.P. Vasco}
\email[]{juan.vasco@epfl.ch}
\affiliation{Institute of Physics, \'Ecole Polytechnique F\'ed\'erale de Lausanne (EPFL), CH-1015 Lausanne, Switzerland}

\author{D. Gerace}
\affiliation{Dipartimento di Fisica, Universit\`a di Pavia, via Bassi 6, I-27100 Pavia, Italy}

\author{V. Savona}
\affiliation{Institute of Physics, \'Ecole Polytechnique F\'ed\'erale de Lausanne (EPFL), CH-1015 Lausanne, Switzerland}



\begin{abstract}
Photonic resonators allowing to confine the electromagnetic field in ultra-small volumes and with long decay times are crucial to a number of applications requiring enhanced nonlinear effects. For applications to integrated photonic devices on chip, compactness and optimized in-plane transmission become relevant figures of merit as well. Here we optimize an encapsulated Si/SiO$_2$ photonic crystal nanobeam cavity at telecom wavelengths by means of a global optimization procedure, where only the first few holes surrounding the cavity are varied to decrease its radiative losses. This strategy allows to achieve close to 10 million intrinsic quality factor, sub-diffraction limited mode volumes, and in-plane transmission above 65\%, in a structure with a record small footprint of around $8~\mu$m$^2$. We address and quantitatively assess the dependence of the main figures of merit on the nanobeam length and fabrication disorder. Finally, we  theoretically give a realistic estimate of the single-photon nonlinearity in such a device, which holds promise for prospective experiments in low-power nonlinear and quantum photonics.
\end{abstract}


\maketitle 

\section{Introduction}

The search for ultra-high $Q/V$ photonic crystal (PC) cavities has attracted very much attention in the field of nanophotonics during the last two decades \cite{vahala}. The diffraction limited confinement enabled by PCs (with mode volumes $V$ of the order of the cubic wavelength in the medium) and the possibility of achieving high quality factors $Q$, have led to several applications in the context of light-matter interaction \cite{deppe,vuckovic3,imamoglu,noda3,vuckovic4,vuckovic2,nomura,vuckovic5,imamoglu2,imamoglu3} and enhanced optical nonlinearities \cite{vuckovic6,wong,notomi3,galli,notomi4,faolain,noda4,momchil3}. One-dimensional PC nanobeam cavities, in particular, have been widely used as they have a significantly smaller footprint with respect to their two-dimensional PC counterparts \cite{quan2}. Cavity designs for fundamental mode quality (Q) factors in the $10^8-10^9$ range and mode volumes of the order of $(\lambda/n)^3$ have been proposed for free-standing nanobeam cavities \cite{notomi5,notomi6}. Moreover, the efforts for increasing $Q/V$ in nanobeam cavities have been mainly focused on silicon photonics at telecom wavelengths \cite{picard,delarue,deotare,dario1,dario4,flayac1,flayac3,witmer,flayac2}, allowing the integration with optoelectronic devices in a single CMOS chip \cite{wang}. 

As far as working performance is concerned, nanobeam systems in free-standing membranes (air-bridge) may be affected by mechanical instabilities and environmental changes, thus making integration challenging. This problem has been solved by means of SiO$_2$ encapsulation \cite{bazin}, with the added benefit that it naturally improves the thermal resistance, as SiO$_2$ provides a better heat sink than air \cite{raineri}, and it mitigates loss channels related to etching of air holes and introduction of leaky surface states in the silicon \cite{borselli}. However, due to the reduction of the contrast between the refractive indices in the system ($n_{{\rm SiO}_2}=1.44$ at telecom), high $Q$ factors are more difficult to achieve, even theoretically, as index-guided confinement becomes less effective. Nevertheless, optimization of Si/SiO$_2$ nanobeam cavity Q-factors has been achieved through a smooth variation of the cavity edges, such that the confined electromagnetic mode follows a Gaussian envelope function \cite{noda,noda5}. This technique, known as the \textit{gentle confinement}, effectively reduces the leaky components of the cavity mode out-of-plane, and it has been exploited to obtain theoretical $Q$ factors of up to $10^7$ for encapsulated nanobeams \cite{quan1}. While the gentle confinement has shown to be very effective to increase the $Q$ factor of nanobeam cavities, the main drawback arises from the large number of holes addressed by the optimization, resulting in very long structures which may be highly sensitive to disorder effects \cite{quan2,quan1}. 

In this work we take a different route, in which only the first few holes surrounding the cavity are varied to decrease radiative losses. This technique has been extremely successful in two-dimensional PC cavities where outstanding figures of merit have been realized \cite{momchil4,momchil5,momchil1,momchil6,momchil7}. Specifically, we use a global optimization strategy combined with first principle numerical simulations in order to maximize the cavity quality factor $Q_c$ at telecom wavelengths, in the space defined by the nanobeam optimization parameters. The optimized cavity designs have quality factors of up to 8 millions, diffraction limited volumes, transmission exceeding 50\% and very small foot print. Additionally, we study the effect of disorder on $Q_c$ and find that an average quality factor of the order of  one million is still achievable when considering state-of-the-art tolerances in sample fabrication techniques. The cavity designs presented in this work are of particular interest for applications in classical and quantum silicon photonics in integrated chips, where the enhancement of linear and non-linear interactions play a fundamental role in the device functionality. To this end, we finally give a realistic estimate of the single-photon nonlinearity to loss rate ratio, which shows the great potential of these systems for quantum photonics experiments in an all-silicon platform.

The paper is organized as follows. In Section~\ref{sec1}, we present the nanobeam cavity optimization and compute the main figures of merit of the optimal designs. In Section~\ref{sec2}, we study the dependence of the total quality factor $Q$ and transmission on the number of holes on each side of the cavity. The effects of disorder are studied in Section~\ref{sec3} and we estimate the  single-photon nonlinear coupling in Section~\ref{sec4}. Finally, the main conclusions of the work are presented in Section~\ref{sec5}.

\begin{figure*}
\includegraphics[width=1\textwidth]{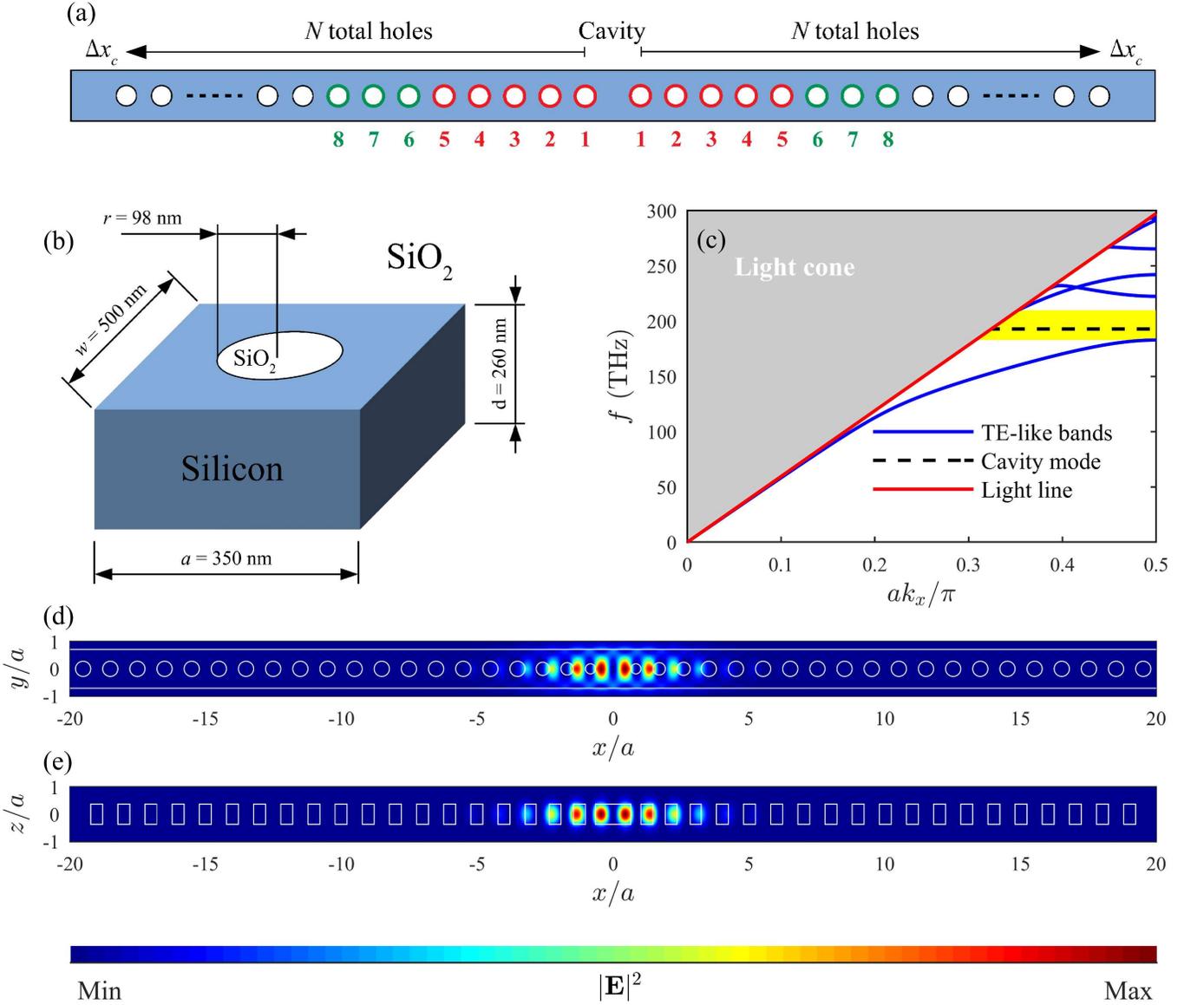}
\caption{(a) Schematic representation of the nanobeam cavity with a total of $N$ holes on each side, and lattice offset $\Delta x_c$. The red and green holes are allowed to vary (size and position), as well as $\Delta x_c$ while preserving the mirror symmetry with respect to the center of the cavity. (b) Parameters of the nanobeam unit cell. (c) TE-like band structure of (b) in the projected Brillouin zone, where the dashed horizontal line represents the non-optimized cavity mode with $f=193$ THz and $Q_c=3.1\times10^4$. The band gap and light cone regions are highlighted in yellow and gray, respectively. (d) Electric field intensity profile $|\mathbf{E}|^2$ of the cavity mode at $z=0$. (e) Same as (d) for $y=0$.}\label{fig1}
\end{figure*}

\section{Nanobeam cavity design optimization}\label{sec1}

We show in Fig.~\ref{fig1}(a) the schematic representation of the nanobeam structure optimized in the present work. The cavity is created by increasing the distance between two adjacent holes by introducing a lattice offset $\Delta x_c$. A total number of $N$ holes is considered on each side of the cavity, and the first 8 ones are allowed to vary (position and size) to optimize the $Q_c$ factor of the fundamental cavity mode. Since we preserve the mirror symmetry with respect to the center of the nanobeam, only the $x$ coordinate and radius $r$ of the hole $i$ (for $1\leq i\leq8$) are changed, i.e., $x'_i=x_i+dx_i$ and $r'_i=r_i+dr_i$, where positive (negative) $dx$ means outward (inward) displacement. We adopt the structure parameters originally reported in Ref.~\onlinecite{dario1} for an asymmetric silicon-on-insulator (SOI) nanobeam cavity structure, i.e. lattice parameter $a=350$~nm, hole radii $r=98$~nm, thickness $d=260$~nm and width $w=500$~nm leading to a resonant mode at telecom wavelengths. 
These parameters are shown in Fig.~\ref{fig1}(b), where the silicon structure is assumed to be completely encapsulated in SiO$_2$, at difference with Ref.~\onlinecite{dario1}. Moreover, our starting cavity design is taken from the same work and it is given by the following set $(dx_1,dx_2,dx_3,dr_1,dr_2,dr_3,\Delta x_c)=(110,60,25,-33,-18,-13,7.5)$~nm, which yields a three-dimensional Finite-Difference Time-Domain (3D-FDTD)\cite{lumerical} fundamental mode quality factor of $Q_c=3.1\times10^4$ at $f=193$~THz. In Fig.~\ref{fig1}(c) we show the projected TE-like band structure of the nanobeam, computed with the MIT photonic bands (MPB) package \cite{steven}, where the yellow region indicates the photonic band gap and the cavity frequency is represented by the horizontal black dashed line. The corresponding intensity profile of the cavity mode, which displays a node at the center of the structure, is shown in Figs.~\ref{fig1}(d) and \ref{fig1}(e), for $z=0$ and $y=0$, respectively. \\
This basic design thus constitute our starting point to carry out the optimization procedure by introducing small modifications of $\Delta x_c$ and the holes surrounding the cavity. Specifically, we consider two independent cases where 5 and 8 holes respectively are varied, thus setting the dimension of the parameter space to 11 and 17, respectively. We use the particles swarm algorithm to carry out the global optimization with $Q_c$ as the objective function, and first-principle FDTD simulations\cite{lumerical} to evaluate $Q_c$. We find optimal cavity quality factors of $7.8\times10^6$ and $8.1\times10^6$ when varying 5 and 8 holes, respectively, at a resonance frequency of $f=194.6$~THz, corresponding to an improvement of two orders of magnitude with respect to the non-optimized design. These quality factors are achieved with a structure as short as $N=20$ (see Fig.~\ref{fig3}), and they are one order of magnitude larger than the ones found in gentle confinement designs for the same sample length in free-standing nanobeams \cite{quan2}. The main figures of merit of our optimized cavities are shown in Table~\ref{tab1}, where the linear mode volume is defined as
\begin{equation}\label{Vl}
 V_l=\frac{\int \epsilon(\mathbf{r})|\mathbf{E}(\mathbf{r})|^2d\mathbf{r}}{\mbox{Max}\left\{\epsilon(\mathbf{r})|\mathbf{E}(\mathbf{r})|^2 \right\}},
\end{equation}
and the non-linear one as \cite{painter}
\begin{equation}\label{Vnl}
 V_{nl}=\frac{\left[\int \epsilon(\mathbf{r})|\mathbf{E}(\mathbf{r})|^2d\mathbf{r}\right]^2}{\int \epsilon^2(\mathbf{r})|\mathbf{E}(\mathbf{r})|^4d\mathbf{r}}.
\end{equation}

\begin{table*}
\caption{\label{tab1} Summary of the linear and non-linear figures of merit for the two optimized nanobeam cavities studied in this work. The ratio $Q_c/V_l$ is relevant to linear phenomena while the $Q_c^2/V^2_{nl}$ is mostly employed for non-linear applications.}
\begin{ruledtabular}
\begin{tabular}{ccccccc}
Varying holes & $f$ (Thz) & $Q_c$ & $V_l$ $(\lambda/n_{\rm Si})^3$ & $V_{nl}$ $(\lambda/n_{\rm Si})^3$ & $Q_c/V_l$ $(n_{\rm Si}/\lambda)^3$ & $Q_c^2/V^2_{nl}$ $(n_{\rm Si}/\lambda)^6$ \\
\hline
5 & 194.6 & $7.8\times10^6$ & 0.38 & 1.92 & $2.05\times10^7$ & $1.65\times10^{13}$ \\
8 & 194.6 & $8.1\times10^6$ & 0.38 & 1.92 & $2.13\times10^7$ & $1.78\times10^{13}$\\
\end{tabular}
\end{ruledtabular}
\end{table*}

\noindent In Eqs.~(\ref{Vl}) and (\ref{Vnl}), $\epsilon(\mathbf{r})$ is the dielectric function of the system and the integration is carried out over the whole computational cell volume. The record-small mode volumes reported in Table~\ref{tab1} for our encapsulated nanobeam cavity optimal designs lead to a $Q_c/V_l$ and $Q_c^2/V_{nl}^2$ enhancement factors  in the $10^7 (n_{\rm Si}/\lambda)^3$  and $10^{13} (n_{\rm Si}/\lambda)^6$ range, respectively, highlighting the potential of these photonic structures for linear and non-linear applications in ultra compact devices. These record values exceed those previously obtained through gentle confinement with $N=40$, i.e. a twice-as-long structure \cite{quan1}. 
The structure parameters obtained for the optimized designs after the particle swarm algorithm are reported in Appendix \ref{A1}, both for the 5 and 8 holes variation. Finally, to visualize the effect of the optimization we have reported in Appendix \ref{A2} also the Fourier transforms of the cavity mode $E_y$ components  for the 5 and 8 holes cases, respectively, highlighting the suppression of the radiative contributions inside the light cone, at the origin of the $Q_c$ enhancement \cite{noda,dario2}. 

\section{Transmission}\label{sec2}

The transmission of the nanobeam can be estimated using the temporal coupled-mode theory presented in Ref.~\onlinecite{joannopoulos}, where the cavity, with quality factor $Q_c$, is assumed to be weakly coupled to the ridge waveguide, leading to an effective $Q_w$ that represents the finite lifetime of the cavity-waveguide coupling. Thus, the total $Q$ of the system can be written as
\begin{equation}\label{Qt}
 \frac{1}{Q}=\frac{1}{Q_c}+\frac{1}{Q_w},
\end{equation}
and the transmission at the cavity resonance takes the simple form 
\begin{equation}\label{T}
T=\frac{Q^2}{Q^2_w}.
\end{equation}

\begin{figure}[t!]
\includegraphics[width=0.48\textwidth]{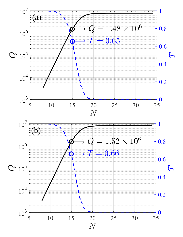}
\caption{ (a) Total quality factor $Q$ (left continuous-black) and transmission $T$ (right dashed-blue) as a function of the total number of holes at each side of the cavity optimized with 5 varying holes. The case with $N=15$ is highlighted as a good compromise with large $Q$ and $T$ while keeping a short length nanobeam. (b) Same as (a) for the optimized cavity with 8 varying holes.}\label{fig3}
\end{figure}

\noindent The behavior of $Q$ and $T$ as a function of the number of holes $N$ is shown in Figs.~\ref{fig3}(a)~and~\ref{fig3}(b) for the 5 and 8 holes optimizations, respectively. The total quality factor of the system displays an exponential increase with $N$, until it eventually starts to saturate to the $Q_c$ value (around $N=20$), corresponding to the limit where the cavity and the waveguide are not coupled anymore, i.e., infinite $Q_w$ and zero transmission. On the contrary, for a decreasing number of holes, $T$ increases and approaches 1 for the smallest sample studied ($N=8$) at the expense of the total $Q$. Optimizing the transmission implies compromising between high $Q$ and high $T$ in order to efficiently inject and extract light from the system, while keeping losses low. We set the optimal transmission at $N=15$, where $Q\sim1.5$ million with transmissions above 60\%. Notice that when considering a gentle confinement geometry in free-standing nanobeams \cite{quan2}, a minimum of $N=15$ should be considered, which leads to $T$ values above 90\%, but $Q$ factors below $10^5$. The present approach allows to keep outstanding figures of merit in encapsulated and very compact structures.

\section{Disorder effects}\label{sec3}

\begin{figure}
\includegraphics[width=0.45\textwidth]{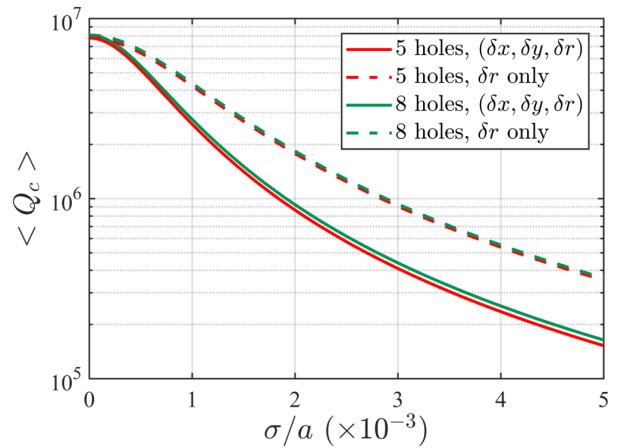}
\caption{ Averaged cavity quality factor $<Q_c>$ over 100 disorder realizations as a function of the disorder parameter for the optimized cavity with 5 (red) and 8 (green) varying holes. Continuous curves correspond to disorder in both, hole position and size, while dashed curves correspond to disorder in the hole size only.}\label{fig4}
\end{figure}

When dealing with realistic samples, disorder is always present, originating from unavoidable imperfections introduced at the fabrication stage. This effect is commonly modeled by assuming non-correlated Gaussian fluctuations $\delta$ in either the position, i.e.,  $(x,y)\rightarrow(x+\delta x,y+\delta y)$, or the size, i.e.,  $r\rightarrow r+\delta r$, of all the holes in the whole photonic structure, where the standard deviation $\sigma$ of the Gaussian probability distribution is taken as the disorder parameter \cite{dario3,momchil2,vasco1,vasco2}. We carried out such an analysis in our nanobeam designs by computing 100 independent disorder realizations for each $\sigma$ value, to obtain the average quality factor of the cavity modes $\langle Q_c\rangle$. Results are shown in Fig.~\ref{fig4} where we plot $\langle Q_c\rangle$ as a function of $\sigma/a$ for the cavities with 5 (red curves) and 8 (green curves) optimized holes. Two independent cases are considered, one in which position and size of all holes are varied at the same time (continuous curves), and another in which only the size of them are fluctuating (dashed curves). Most often, the latter is believed to be the dominant source of fabrication imperfection due to the etching process. We specifically compute $\langle Q_c\rangle$ for $\sigma=0.001a$ and $\sigma=0.005a$, respectively, and use the linear relation between $\langle Q_c\rangle^{-1}$ and $\sigma^2$ to interpolate the corresponding curve \cite{momchil2}. We notice that typical state-of-the-art tolerances in silicon PCs range between $\sigma=0.002a$ and $\sigma=0.003a$ \cite{noda2,mohamed2}, leading to disorder-induced losses quality factors of $Q_d\simeq2.5\times10^6$ and $Q_d\simeq1.1\times10^6$, respectively, for dominant size disorder and an averaged $Q_c$ in the million range as shown in the Figure. The value of $Q_d$ is estimated by means of the simple relation $1/\langle Q_c\rangle\simeq1/Q_c+1/Q_d$ (see Ref.~\onlinecite{momchil2}). We also notice that $Q_c$ factors in the half million range have been previously measured in encapsulated Si/SiO$_2$ nanobeams \cite{quan1}, but in much longer nanobeam cavities ($N=40$) where the role of disorder might be more relevant.

Our results hold promise that fully encapsulated nanobeam cavities with a million quality factor and a large in-plane transmission may be realized with state-of-art SOI technology.

\section{Single-photon nonlinear coupling}\label{sec4}
 
Integrated photonic devices with the characteristics fulfilled by our optimized encapsulated nanobeam cavities may be key to a number of applications requiring enhanced on-chip nonlinearities, such as optical routing, neuromorphic computing, or entangled photon pair generation by spontaneous four-wave mixing. We give here an estimate of the single-photon nonlinearity, i.e. the interaction energy between two photons simultaneously present in the resonator and induced by a given nonlinear process. This is a key quantity to consider for low-power nonlinear optics applications, or more recent proposals for quantum photonics experiments in passive materials (i.e., not exploiting intrinsic resonances of the medium). For silicon, it is relevant to calculate this contribution when it is induced by a third-order nonlinear susceptibility of the material, quantified by the $\chi^{(3)}$ tensor elements. 
For the case of passive nonlinear resonator, the nonlinear energy shift for a resonant mode at frequency $f$ is given by the relation \cite{dario4,ferretti} 
\begin{equation}\label{shift_kerr}
\begin{split}
U_{\mathrm{nl}} =& \frac{3(h f)^2 } {4\varepsilon_0} \int \chi^{(3)}(\mathbf{r})|\mathbf{E}(\mathbf{r})|^4 \mathrm{d}\mathbf{r} \\
  \simeq  & \frac{3(h f)^2 } {4\varepsilon_0 V_{\mathrm{nl}}} \frac{\overline{\chi}^{(3)}}{{\varepsilon}^2}
\end{split}
\end{equation}
where the approximate expression is given in terms of the nonlinear mode volume for third-order nonlinear processes, Eq.~(\ref{Vnl}), and $\overline{\chi}^{(3)}$ represents a weighted value of the nonlinear susceptibility that takes into account its spatial dependence. To give quantitative estimates, we assume constant values for the real part of the nonlinear susceptibility and relative dielectric permittivity that are known for silicon at telecom wavelengths \cite{hon}, i.e. ${\chi}^{(3)}=0.25\times 10^{-18}$ m$^2$/V$^2$ and ${\varepsilon}_r =12.09$, respectively, and neglect nonlinear contributions from the silica. For our optimized cavities, we numerically calculated the integral in Eq.~(\ref{shift_kerr}),  obtaining the value $U_{\mathrm{nl}}\simeq 1.6 \times 10^{-4}$ $\mu$eV for structures with both 5 and 8 modified holes. From results of Fig.~\ref{fig3}, and taking into account that $Q_d\simeq 2.5\times 10^6$ with a size disorder of $\sigma/a=0.002$ ($\sigma\sim 0.7$ nm for our current design), we can realistically achieve a $Q \sim 10^6$ ($1/Q \rightarrow 1/Q+1/Q_d$) at telecom energies (0.8 eV) with a sample length of $N=15$. This gives a theoretical ratio between nonlinearity and intrinsic cavity loss rate in the order of $U_{\mathrm{nl}} Q / h f \sim 2 \times 10^{-4}$. While such value makes it very challenging to directly probe single-photon nonlinearities with these devices \cite{dario4}, it is a very promising estimate for quantum photonics experiments relying, e.g., on quantum interference effects \cite{flayac3,flayac4}.

 \section{Conclusions}\label{sec5}
 
 We have designed an encapsulated Si/SiO$_2$ photonic crystal nanobeam cavity by means of a global optimization strategy, where only the closest holes to the cavity are varied to increase the quality factor of the system. Differently from the commonly adopted \textit{gentle confinement} mechanism, our approach results in a very small footprint structure of around $8~\mu$m$^2$ with total $Q$ of several millions, small linear mode volume in the 0.4 $(\lambda/n)^3$ regime and transmission above 50\%, thus setting record figures of merit for such ultra-compact photonic device. We have studied the effects of intrinsic disorder on the quality factor of the nanobeam cavity and found that, when considering typical tolerances achieved in modern sample fabrication techniques, it remains in the million range, which still correspond to an outstanding result given the short length of the structure. Our nanobeam designs are of special relevance for applications in integrated photonics where extremely large $Q_c/V_l$ and $Q_c^2/V_{nl}^2$ factors are required for enhanced optical nonlinearities. To corroborate our findings, we have estimated a realistic single-photon to loss rate ratio of $\sim2\times10^{-4}$ in our best compact devices, in which the role of disorder and fabrication imperfections is also taken into account. The latter is among the highest values reported in the literature for such figure of merit, and will motivate realizing these devices in quantum photonic experiments on chip. 
 
 \section{Acknowledgments}
 Several useful discussions with Marco Clementi, Thomas Fromherz, Matteo Galli, and Momchil Minkov are gratefully acknowledged. DG acknowledges financial support from the EU H2020 QuantERA ERA-NET Co-fund in Quantum Technologies project CUSPIDOR, co-funded by the Italian Ministry of Education, University and Research (MIUR).

 \appendix
 
 \section{Optimized cavity parameters} \label{A1}

Here we report the structural parameters obtained from our particle swarm algorithm, for the optimal designs when 5 and 8 are varied.
The modified holes' parameters are reported in Table~\ref{tab2}.  

  \begin{table}[h!]
\caption{\label{tab2} Optimal parameters for the cavities with 5 and 8 varying holes found by the global optimization.}
\begin{ruledtabular}
\begin{tabular}{ccccccccc}
& hole 1 & hole 2 & hole 3 & hole 4 & hole 5 & hole 6 & hole 7 & hole 8 \\
\hline
&  \multicolumn{7}{c}{\textit{Optimization with 5 varying holes $\Delta x_c=29.8$} nm}  &\\
$dx$ (nm) & 36.3 & 6.5 & 4.2 & 7.8 & 11.2 & 0 & 0 & 0\\
$dr$ (nm) & -44.7 & -30.6 & -12.5 & 5.2 & 11.8 & 0 & 0 & 0\\
\hline
& \multicolumn{7}{c}{\textit{Optimization with 8 varying holes $\Delta x_c=29.7$} nm}  &\\
$dx$ (nm) & 36.2 & 6.7 & 4.2 & 7.8 & 11.1 & 0.1 & 0 & 0\\
$dr$ (nm) & -44.6 & -30.5 & -12.5 & 5.1 & 11.8 & 0.3 & 1.4 & 0\\
\end{tabular}
\end{ruledtabular}
\end{table}

 \begin{figure}[h!]
 \includegraphics[width=0.45\textwidth]{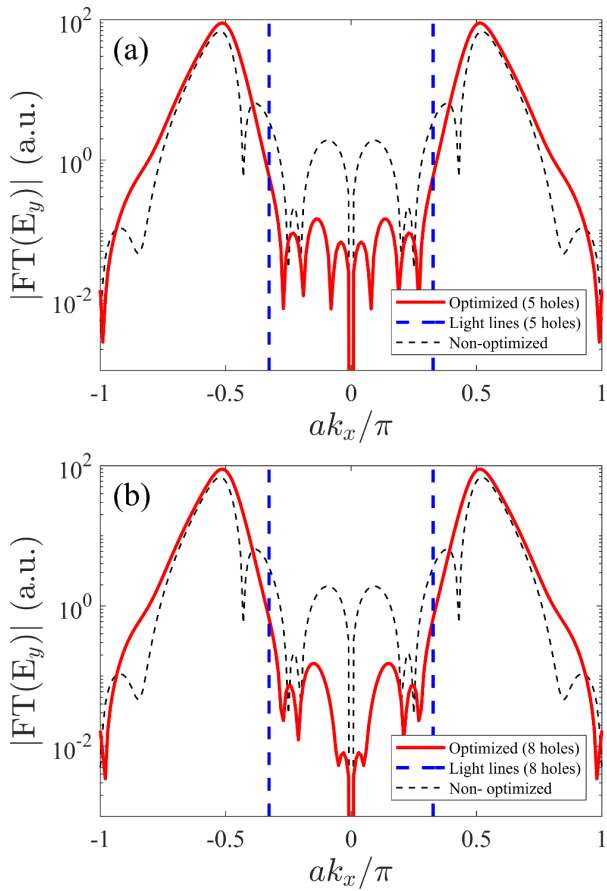}
 \caption{\label{fig2} (a) Projected Fourier transform of the E$_y$ component along the $x$ direction at $z=y=0$ for the cavity optimized with 5 varying holes (continuous red), superimposed over the corresponding curve of the non-optimized cavity (dashed black). The light lines for the optimized cavity are shown in dashed blue. (b) same as (a) for the cavity optimized with 8 varying holes.}
 \end{figure}

\section{Fourier transform of near-field components} \label{A2}

To visualize the effect of optimization, we plot in Figs.~\ref{fig2}(a) and \ref{fig2}(b), the far-field projection of the cavity mode along the $k_x$ direction at $k_y=0$, for the 5 and 8 holes cases, respectively. The vertical blue dashed curves correspond to the points where the cavity frequency crosses the light line, while the black dashed curves are the corresponding far-field projections of the non-optimized cavity. The far-field is obtained through the Fourier transform of the near-field computed in a $xy$ plane localized 70~nm above the nanobeam surface, and because of the even symmetry of $E_y$ with respect to $y$ coordinate, it is dominated by the $E_y$ near-field component when $k_y=0$ (see Ref.~\onlinecite{vuckovic}). Figure~\ref{fig2} shows the suppression of the radiative contributions to the far-field inside the light cone of the structure for the optimal cavity designs, which results in the $Q_c$ enhancement \cite{noda,dario2}.    
 
\newpage


%

\end{document}